# Moiré, Euler, and self-similarity – the lattice parameters of twisted hexagonal crystals

M. Feuerbacher
*Ernst Ruska-Centre for Microscopy and Spectroscopy with Electrons and Peter Grünberg Institute, Forschungszentrum Jülich GmbH, 52425 Jülich, Germany*

A real-space approach for the calculation of the Moiré lattice parameters for superstructures formed by a set of rotated hexagonal 2D crystals such as graphene or transition-metal dichalcogenides, is presented. Apparent Moiré lattices continuously form for all rotation angles, and their lattice parameter in a good approximation follows a hyperbolical angle dependence. Moiré crystals, i.e. Moiré lattices decorated with a basis, require more crucial assessment of the commensurabilities and lead to discrete solutions and a non-continuous angle dependence of the Moiré-crystal lattice parameter. In particular, this lattice parameter critically depends on the rotation angle, and continuous variation of the angle can lead to apparently erratic changes of the lattice parameter. The solutions form a highly complex pattern, which reflects number-theoretical relations between formation parameters of the Moiré crystal. The analysis also provides insight into the special case of a 30° rotation of the constituting lattices, for which a dodecagonal quasicrystalline structure forms.

## 1. Introduction

In 2018 Cao et al. [1] demonstrated that stacked graphene layers with relative rotations can have drastically different properties than their regularly aligned counterparts. They observed that a relative rotation of about 1.1° leads to superconductivity in double-layer graphene. This corresponds to one of the "magic angles" previously predicted by Bistritzer and MacDonald [2], who calculated the Band structure of twisted double-layer graphene and found that the narrowing of bands at small angles is non-continuous, and at 1.05° and other distinct angles, the Dirac-point velocity vanishes. These publications, and several others dealing with rotation-controlled band-structure modifications in double-layer systems (see e.g. [3 – 7]) opened up the field of "twistronics" which currently attracts considerable scientific interest. Also at large rotation angles, twisted bilayers bear striking effects, such as the formation of a dodecagonal quasicrystalline phase at a rotation angle of 30°.

In order to fully understand the physical effects in twistronics and to allow precise device design, it is imperative to understand the relation of the relative rotation angle of the constituting layers and the resulting structure of the Moiré crystal. This is particularly important due to the fact that both the physical properties – viz. the occurrence of distinct "magic angles" – as well as the lattice parameter of the Moiré crystal, and therewith the local order and the length scale of its modulation, are critically dependent on the rotation angle. Recently the field evolved, and besides graphene bilayers now also Dichalcogenides such as MoS2 and WS2 [8,9] as well as heterogeneous systems, e.g. stacked BN and graphene layers [10] or MoS2 on graphite [8], are considered.

In this paper a straightforward and presumption-free framework for the direct calculation of the angle dependence of the Moiré crystal lattice parameter is provided. The paper has two aims: First, to establish a set of equations that describe the Moiré lattice parameter as a function of the rotation angle of the constituting lattices that can straightforwardly be used for spreadsheet calculation or be implemented in compact computer code for everyday use e.g. in the design of twisted-bilayer devices or the interpretation of high-resolution transmission electron micrographs of twisted-bilayers. For example, the equations can be used to identify angles of particularly critical angle-dependence, or for back-calculation of the rotation angle by measurement of lattice parameters of actual devices in the transmission electron microscope for quality control or reproducibility checks. Second, to point out the salient relations between the solution pattern of Moiré-crystal lattice parameters and mathematical number theory, which may imply a corresponding relation between number theory and the actual properties of twisted bilayer devices.

## 2. Phenomenology

When two hexagonal lattices are subjected to relative rotation (Fig. 1) Moiré patterns occur. These patterns, at angles below and above 30°, have the same rotational symmetry and consist of area-filling rhombuses as the constituting hexagonal lattices but they are rotated with respect to the latter, and the periodicity length, in the following referred to as the Moiré lattice parameter, is larger.

Starting at low angles of relative rotation the pattern develops in the form of broad zones of high and low net-plane density. These zones are broad at low angles, because the net planes are almost parallel in the constituting lattices. The rotation center appears bright because at the common point of rotation the lattice planes converge and do not fill up the space in between. Further bright zones appear on a line through the origin about perpendicular to the a-axis of one of the lattices, in the following referred to as the reference lattice, and about perpendicular to the b-axis of the other lattice, referred to as the rotated lattice. The angle between these lines, in the following referred to as Moiré lines, is thus 60°, corresponding to the narrow angle in the rhombic cell used for the description of a hexagonal structure. The symmetry of the underlying lattices demands that further Moiré lines parallel to the center-crossing lines occur, which consequently form a rhombic Moiré pattern.

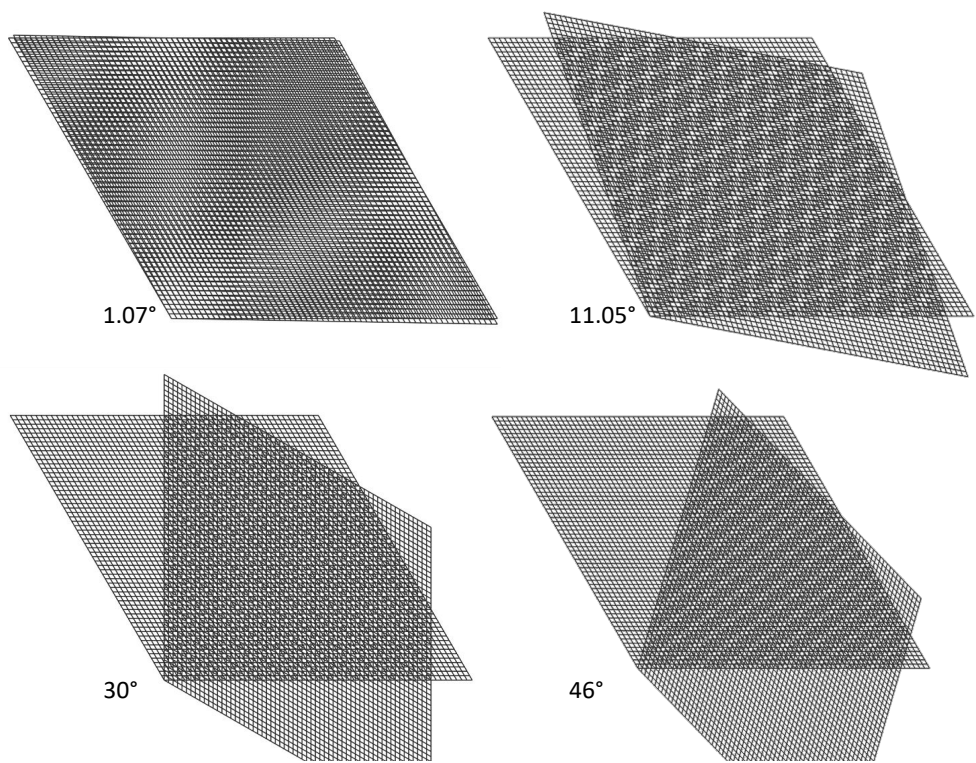

Fig. 1: Moiré patterns at different angles and dodecagonal quasicrystal at 30° formed by relative rotation of two hexagonal lattices.

As the rotation proceeds the bright low-density zones move towards the rotation center and new bright zones occur far out and move towards the axes of the constituting lattices, thus forming a Moiré lattice of increasingly smaller rhombuses. Since the angle of the Moiré line-sets, with respect to the normal of the a- and b- axis of the reference and rotated lattice, respectively, increases by half the rotation angle of the underlying lattices, the rhombs of the Moiré pattern are always oriented with the narrow angle towards the rotation center.

Upon further rotation, lattice parameter of the Moiré pattern decreases continuously, until, at angles approaching 30°, which corresponds to half the periodic angle, it approaches the length scale of the underlying lattices. Around 30° no discernible Moiré pattern is visible, but highly complex structures form. This is the case in a range of about 8° below and above 30°, I.e. from about 22° to 38°. At precisely 30° a dodecagonal quasicrystalline pattern occurs. At angles larger than 38°, a Moiré pattern becomes visible again, developing from smaller to larger Moiré lattices with increasing rotation angle. The angle of the Moiré line-sets with respect to the normal of the a- and b- axis of the reference and rotated lattice, respectively, is now given by half the rotation angle of the underlying lattices plus 30°, and therefore the Moiré pattern still appears as lattice of rhombuses with the narrow edge oriented towards the rotation center.

The lattice parameter of the Moiré lattice can be determined by calculating the distance from the rotation center to the center of the first bright zone on the center-crossing lines. We will refer to this first bright zone as the first order Moiré point. The second bright zone will be referred to as second order Moiré point, etc. Due to the discrete nature of the lattices, not in all Moiré points the lines of the underlying lattices cross perfectly in one point. We will refer to Moiré points, in which the lines cross perfectly (or very close) as "clean" Moiré points. For an overview of the nomenclature and coordinate systems used, see Fig. 2.

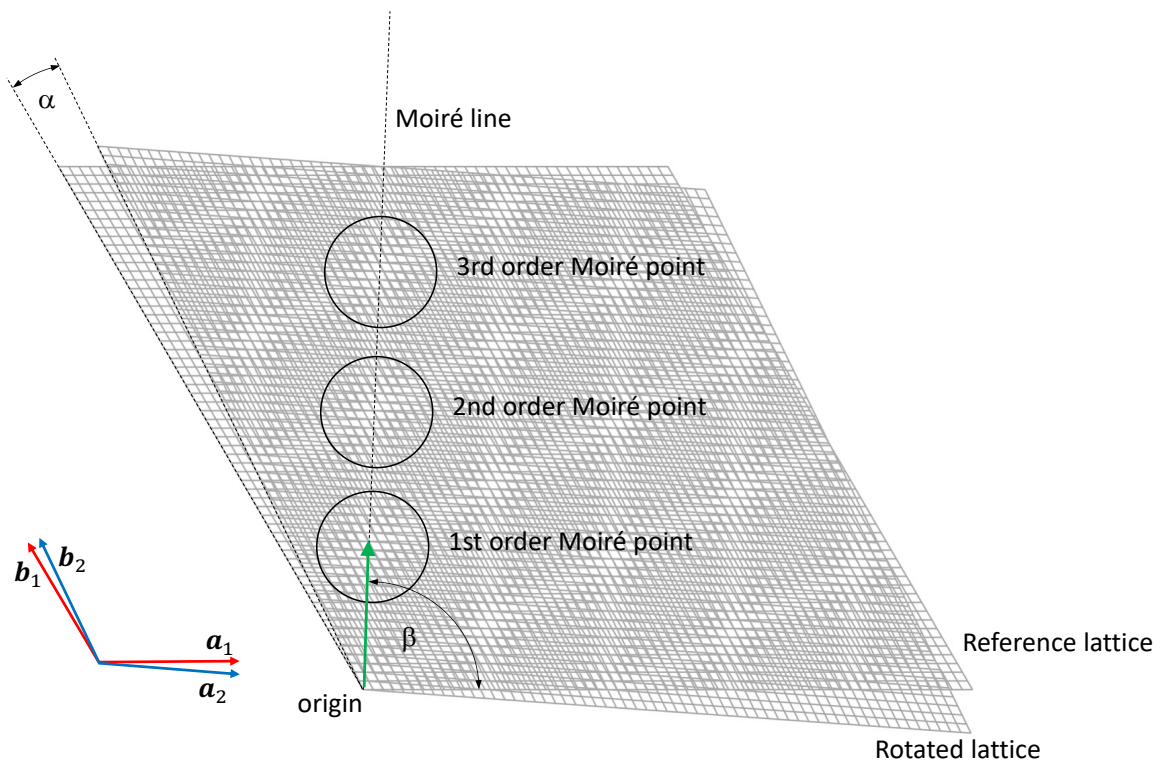

Fig. 2: Coordinate system and nomenclature for reference and rotated hexagonal lattices (here rotated by $\alpha$ = 4.3°). The vector pointing at the first Moiré point is shown in green.

## 3. Calculation of the Moiré lattice parameters

Be $a_1$, $b_1$, the basis of the reference lattice, $a_2$, $b_2$, the basis of the rotated lattice and $\alpha$ the angle of relative rotation. The Moiré lattice resulting from the rotation has basis vectors $a_M$, $b_M$. The direction for $a_M$ at $\alpha = 0$ is chosen perpendicular to $a_1$, which implies that it can be represented by a positive linear combination of $a_1$, $b_1$, and $a_2$, $b_2$.

The angle between $a_1$ and $a_M$ will be referred to as $\beta$, which is thus 90° for $\alpha = 0$, and, as written above

$$\beta = 90° - \alpha/2 \qquad \text{for } \alpha < 30°, \tag{1}$$

$$\beta = 60° - \alpha/2 \qquad \text{for } \alpha > 30°. \tag{2}$$

In order to find the Moiré lattice parameter, we thus have to find the first point on the Moiré line, which is common to both the reference lattice and the rotated lattice. Generally, we can express a point of the Moiré lattice in terms of the reference lattice or the rotated lattice, respectively, as

$$a_M = n\, a_1 + m\, b_1 \tag{3}$$

$$a'_M = n'\, a_2 + m'\, b_2, \tag{4}$$

where $n$, $m$, $n'$ and $m'$ are integers. These, however, cannot be any vector in the lattices – our choices above limit down the directional range of the vectors and therewith establish relations between the coefficients $n$, $m$, $n'$ and $m'$. Fixing the direction perpendicular to $a_1$ at $\alpha = 0$ immediately leads to the condition $m' = m$. Furthermore the direction of $a_M$ for $\alpha \neq 0$ relates the coefficients $n$ and $m$, in particular we find

$$2n - m = p \qquad \text{for } \alpha < 30° \tag{5}$$

and

$$2m - n = p \qquad \text{for } \alpha > 30°, \tag{6}$$

where $p$ is the order of the Moiré point.

Let us now calculate the angle $\beta$ between the vector $a_M$, expressed in terms of the reference lattice, and the $a_1$ lattice vector, via the scalar product $\cos\beta = \frac{a_1 \cdot a_M}{a_1 a'_M}$

Using the basis vectors of the reference lattice

$$a_1 = \begin{pmatrix} 1 \\ 0 \end{pmatrix}, b_2 = \begin{pmatrix} -\cos(60°) \\ \sin(60°) \end{pmatrix} = \tfrac{1}{2}\begin{pmatrix} -1 \\ \sqrt{3} \end{pmatrix}, \tag{7}$$

we obtain the relation

$$\cos\beta = \frac{n - \frac{m}{2}}{\sqrt{n^2 + m^2 - nm}}. \tag{8}$$

This relation can immediately be used to identify candidates for Moiré lattice points: for a given angle $\beta$ we have to find combinations of integers $n$ and $m$ fulfilling (8) within a defined error margin, which then, using (3) can be used to calculate the corresponding vectors $a_M$ in the basis of the reference lattice.

Let us now express a point of the Moiré lattice $a_M$ in terms of the rotated lattice according to (4), using the basis

$$a_2 = \begin{pmatrix} \cos \alpha \\ -\sin \alpha \end{pmatrix}, b_1 = \begin{pmatrix} \cos(60° + \alpha) \\ -\sin(60° + \alpha) \end{pmatrix} = \frac{1}{2}\begin{pmatrix} \cos \alpha - \sqrt{3} \sin \alpha \\ \sqrt{3} \cos \alpha + \sin \alpha \end{pmatrix}, \quad (9)$$

and again calculate the scalar product of this vector $\cos \beta = \frac{a_1 \cdot a'_M}{a_1 a'_M}$ with the reference lattice. We obtain the relation

$$\cos \beta = \frac{\left(n' - \frac{m}{2}\right)\cos \alpha + \frac{\sqrt{3}}{2} m \sin \alpha}{\sqrt{n'^2 + m^2 - n'm}}. \quad (10)$$

Since both scalar products calculated relate to the $a_1$ vector of the reference lattice, they both describe the same angle. Therefore (8) and (10) can be equated, which gives us a condition for the identification of Moiré points, as lattice points common to both the reference and the rotated lattice.

Because the denominators are independent of the rotation angle, the enumerators and denominators have to be equal independently. Equating the denominators gives us the condition $n^2 + m^2 - nm = n'^2 + m^2 - n'm$. This can be fulfilled only if $n' = m - n$, which establishes another relation between the coefficients.

Equating the enumerators finally gives us a relation between the rotation angle and the coefficient $n$ and $m$:

$$\tan\left(\frac{\alpha}{2}\right) = \frac{1}{\sqrt{3}} \frac{2n-m}{m}. \quad (11)$$

For first order Moiré points, using (5) with $p = 1$, we then obtain

$$m = \frac{1}{\sqrt{3} \tan(\alpha/2)}. \quad (12)$$

For a given rotation angle, we can now use (12) to calculate $m$, then calculate $n$ using (5) or (6) under the condition that both $m$ and $n$ are integers. Then (3) is used to calculate the basis vector $a_M$ of the Moiré lattice and its length, the Moiré lattice parameter.

| α / degrees | n | m | $a_M/a_1$ |
|---|---|---|---|
| 0.21 | 158 | 315 | 272.80 |
| 1.09 | 31 | 61 | 52.83 |
| 2.13 | 16 | 31 | 26.85 |
| 2.87 | 12 | 23 | 19.92 |
| 3.15 | 11 | 21 | 18.19 |
| 5.09 | 7 | 13 | 11.27 |
| 6.01 | 6 | 11 | 9.54 |
| 9.43 | 4 | 7 | 6.08 |
| 13.17 | 3 | 5 | 4.36 |
| 21.79 | 2 | 3 | 2.65 |

Table 1: Integer solutions of eq.(12) and lengths of Moiré lattice vectors.

Table 1 displays integer solutions for Moiré lattice parameters, calculated according to this procedure accepting a deviation of ± 0.005 for *n* and *m* from being integers and sampling the rotation angle at $10^{-5}$ degrees. The full set of solutions between 0.1 and 30 degrees is displayed in Fig. 3.

Note that the last entry involves the indices *n,m* = 2,3, implying *n'* = 1, which is the lowest possible combination of positive integers ≠ 0 and corresponds to an angle of 21.8°. This is in agreement with our previous observation that in the range between 22° and 30° no Moiré lines can be seen.

The length of the Moiré lattice parameter can be directly calculated from the first-order Moiré points. Using (3) and (5) for *p* = 1 we obtain

$$a_M = \frac{1}{2}\sqrt{3m^2 + 1} \,. \tag{13}$$

Combining this result with (12), the angular dependence of the length of the Moiré lattice parameter is obtained as

$$a_M = \frac{1}{2\sin(\alpha/2)} \,. \tag{14}$$

For small angles $\alpha$, the sine can be very well approximated by its argument, which leaves us with the expression $a_M \approx \frac{1}{\alpha}$ ($\alpha$ in radians) or, in degrees, $a_M \approx \frac{180}{\pi \alpha}$.

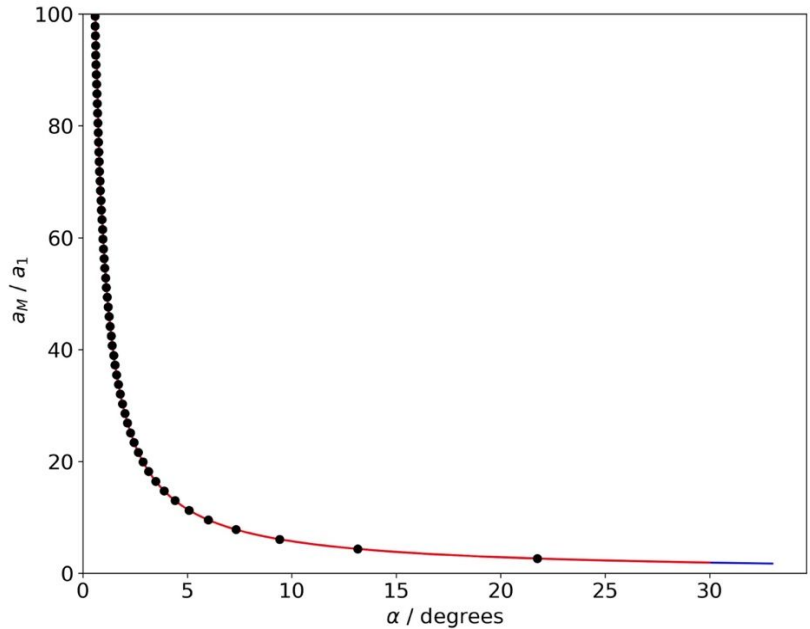

Fig. 3: Moiré lattice parameter (in units of the reference lattice parameter) as a function of $\alpha$. Red: exact solution; blue: approximation $a_M$ = 180/$\pi\alpha$; black: length of actual Moiré-lattice basis vectors for which *m* and *n* are integers (Table 1).

Fig. 3 compares this continuous description with the length of actual basis vectors of the Moiré lattice, for which *m* and *n* are integers. This graph can be understood in the following way: the continuous line according to (14) describes the lattice parameter of all Moiré lattices, as we apparently see them developing in a set of rotated lattices. If we take a closer look, however, we see that for most angles, the first order Moiré points are not clean – the net planes of the reference and the rotated lattice do not meet in a single point, but just get more or less close.

The black circles in Fig. 3, corresponding to the solutions of (12) for which *m* is integer, are the Moiré-lattice parameters of *clean* Moiré points, in which the underlying net planes meet in a single point. The angle dependence of Moiré points is thus continuous, while the angle dependence of *clean* Moiré points is discrete.

For the sake of completeness, let us introduce the b-lattice vector, $b_M$, for the Moiré lattice. We choose the basis such that $b_M$ is perpendicular to the b-axis of the rotated lattice, $b_2$, for $\alpha = 0$ (Fig. 4). This gives us a basis that is not right handed, but naturally suits the resulting Moiré lattice. In particular, the Moiré points corresponding to positive linear combinations of $a_1$, $b_1$, and $a_2$, $b_2$ also have positive indices in terms of the Moiré-lattice basis.

The b-axis vector of the Moiré lattice in terms of the reference lattice is then given by

$$b_M = m\, a_1 + (n-1)\, b_1, \tag{15}$$

which together with (3) forms the basis of the Moiré lattice.

All points of the Moiré lattice $A_M$ can then be expressed using two indices *p* and *q*, where *p* is the previously introduced order of the Moiré points along the *a*-axis, and *q* is the corresponding parameter along the direction of the *b*-axis, as

$$A_M = p\, a_M + q\, b_M. \tag{16}$$

The Moiré lattice for a given rotation angle of the constituting lattices can thus fully be described by the following procedure: for a given rotation angle $\alpha$, use (12) to calculate *m*, then calculate *n* using (5) for *p* = 1. With (3) and (15) the Moiré-lattice basis for the given angle is obtained, which, using the indices *p* and *q*, spans the full Moiré lattice (Fig. 4).

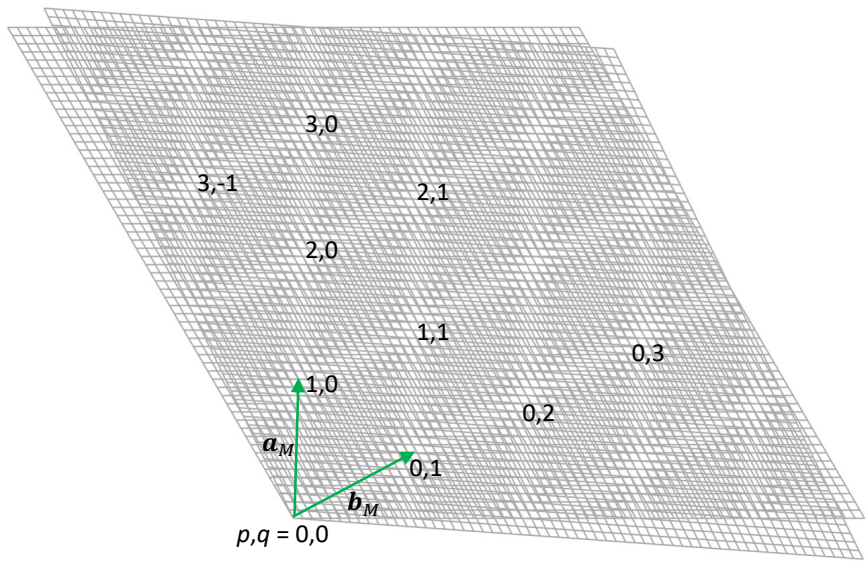

Fig. 4: The Moiré lattice spanned by the basis $a_M$, $b_M$ (green arrows) and its indexing using the coefficients *p* and *q*.

## 4. Higher-order Moiré points and Moiré crystals

Let us now consider Moiré crystals, i.e. we add a basis to the constituting lattices. If we want to consider Moiré crystals we necessarily have to take into account higher-order Moiré points, as can be seen from the following example.

Fig. 5a displays a Moiré lattice at $\alpha = 6.6°$. The Moiré points of order *p* = 0,1,2,3 can clearly be seen. If we look closer, however, we recognize that the first-order Moiré point is not clean (red circle), but the second-order Moiré point is (blue circle). This has a striking effect if we consider crystals rather than mere lattices: in Fig. 5b a Moiré crystal resulting from two Graphene layers, rotated by the same angle of 6.6° is displayed. In the first-order Moiré point (red circle) now a completely different local atomic arrangement than in the second-order point (blue circle) appears. The atomic arrangement in the second-order Moiré point corresponds to that of the origin, and therefore the distance between these constitutes the Moiré lattice parameter for the given rotation angle. The Moiré crystal unit cell, correspondingly, is given by the green rhomb in Fig. 5b, and has twice as long a lattice parameter than a first look on the lattice alone would suggest.

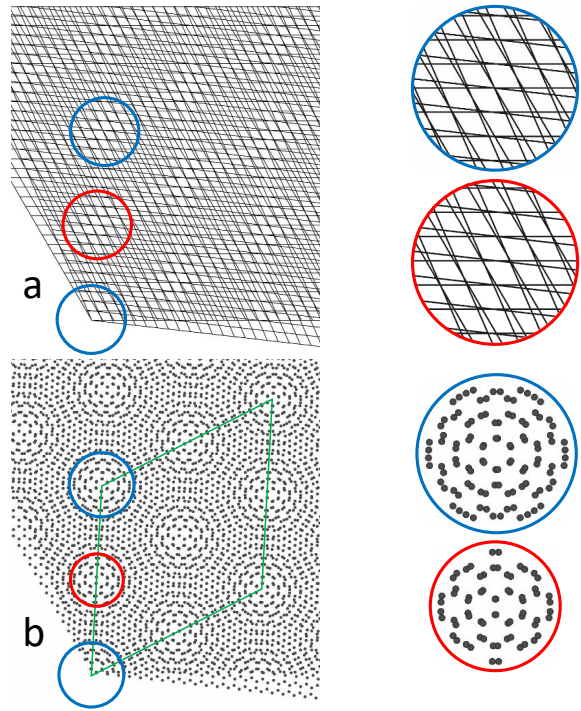

Fig. 5: Moiré lattice (a) and Graphene Moiré crystal at a rotation angle of 6.6 degrees. The circles display the lattice planes and local atomic arrangement at the first-order (red) and second-order (blue) Moiré point.

The example illustrates a general property of Moiré crystals: at a given rotation angle between the constituting lattices, *the lattice parameter of the resulting Moiré crystal is given by the clean Moiré point of lowest order at that angle.*
In order to describe the Moiré crystal we thus have to calculate the higher-order Moiré points, and then figure which of those are clean. For this we combine (11) and (5) which yields

$$\tan\left(\frac{\alpha}{2}\right) = \frac{1}{\sqrt{3}} \frac{p}{m} . \tag{17}$$

The length of the vectors to higher-order Moiré points is calculated directly from (3) and using (5) we obtain

$$a_{M,p} = \frac{1}{2}\sqrt{3m^2 + p^2} , \tag{18}$$

and combining (17) and (18) yields the dependence on the rotation angle $\alpha$

$$a_{M,p} = \frac{p}{2} \frac{1}{sin(\alpha/2)}, \tag{19}$$

for a given order $p$. We will refer to $a_{M,p}$ as the length of the higher-order Moiré lattice vectors, the angle dependence of which is now represented by a discrete family of curves. The curve for $p = 1$ corresponds to the line in Fig. 3.

The lattice parameter of the Moiré crystal for a given angle is determined by finding the solutions of (17) for integers $m$, which, for most angles, will yield more than one solution for different orders $p$. The solutions for 1 to 30°, taking into account the first 30 orders is shown in the Appendix, Fig. A1.

From these solutions, the one with the lowest order corresponds to the Moiré lattice parameter, all higher orders are mere multiples of the latter and can be neglected. With the so determined values for $m$ and $p$, the length of the Moiré lattice parameter is calculated using (18), and the Moiré lattice vectors by (3) and (15) using the value for $n$ obtained through (5).

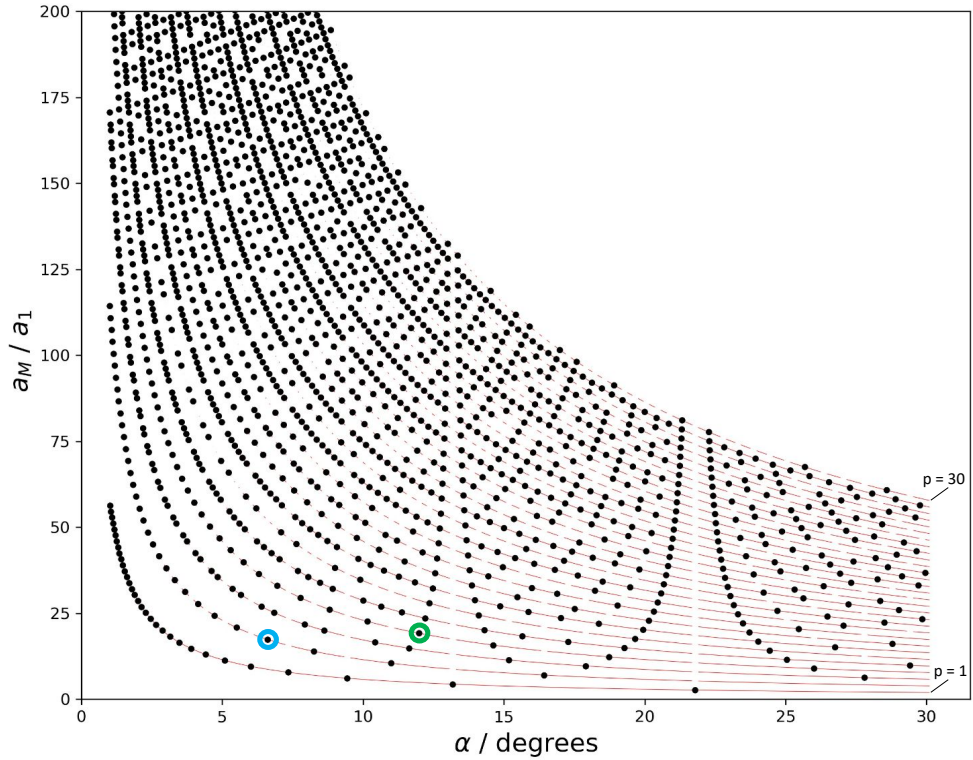

Fig. 6: Moiré crystal lattice parameters (in units of the reference lattice parameter) as a function of $\alpha$ for $p = 1$ to 30. The continuous angle dependence of the Moiré lattice points are shown as guide to the eye. The lines are broken at the positions of higher-order multiples.

Figure 6 displays the so determined angle dependence of the Moiré crystal lattice parameter for angles between 1 and 30° up to the 30[th] order. Also shown is the family of curves (red lines in Fig. 6), representing the continuous solutions of (19). The positions of the higher-order multiples of the Moiré crystal lattice parameters are indicated as breaks in the red curves.

Each dot in the solution pattern Fig. 6 represents a Moiré crystal, which forms at the respective angle with the shown lattice parameter. The green circle, for example, marks a point at the angle 11.98°, which means that here a solution is found for a Moiré crystal with a lattice parameter of 19.16 $a_1$ at a Moiré point of 4[th] order. At this angle, there exists no clean solution at a Moiré point of lower order that would lead to a shorter lattice parameter. A slightly lower rotation angle of 11.64° leads to a Moiré crystal with a lattice parameter of 14.80 $a_1$ (at a 3[th] - order point), and a slightly larger angle of 12.20° to a Moiré crystal with a lattice parameter of 23.52 $a_1$ (at a 5th -order point). The blue circle marks the example shown in Fig. 5, a Moiré crystal at 6.6° at a 2nd order point.

The diagram shows that Moiré crystals exist for all angles, but for some of them, Moiré points of very high order have to be taken into account. These high-order points correspond to very large Moiré lattice parameters that may exceed the size of the flake of sample material, which implies an ultimate upper limit for the relevance of the high orders. For graphene, which has an $a$-lattice parameter of about 2.5 Å, a typical consistent flake size of 1 µm corresponds to solutions of about 80th to 100th order.

The diagram also shows that the Moiré lattice parameter, critically depends on the angle. In some ranges, variation of $\alpha$ leads to small variations of the Moiré lattice parameter, but more often, small angle variations lead to considerable changes in lattice parameter. This is for example the case for the angles for which first-order solutions exist, e.g. at 22°. A small change in angle leads to a change to a large lattice parameter of the highest order considered. Accordingly, the sequence of Moiré crystal lattice parameters upon variation of the rotation angle can be seemingly erratic.

Fig. 7a depicts the evolution of the Moiré lattice parameter for the case of graphene around the magic angle of 1.1° (blue line), taking into account orders of up to 8. Within the small angular range of 0.035° the Moiré crystal lattice parameter takes on various, greatly different values between 130 and 1050 Å. Even if only second- or third-order solutions are taken into account, the lattice parameter can take values varying by a factor of almost 2 or 3, respectively. The figure also shows that at the magic angle of 1.1° no first-order solution exists. The closest solution is of 7th order and corresponds to a lattice parameter of about 910 Å. The closest first-order solution, on the other hand, occurs at an angle of 1.085° and corresponds to a lattice parameter of about 130 Å. This comes close to the value mentioned by Cao et al. [1]. Fig. 7b is a similar depiction of the situation at higher angles, 13° to 20°, taking into account orders of up to 16. The critical angle dependence, which includes changes of the lattice parameter of up to 160 Å upon minor changes of the rotation angle, is clearly seen.

The solution pattern of the Moiré crystal lattice parameters depicted in Fig. 6, is separated into intervals which are defined by the angle positions of the first-order solutions. The most obvious interval in the figure extends from 13.2° to 21.8°, narrower ones extend from 9.4° to 13.2°, from 7.3° to 9.4°, etc., and an incomplete interval extends from 21.8° to higher angles. At the limits of the intervals the solutions apparently diverge, which can be described by a set of pole functions (Appendix A2.2). Note that each interval contains an identical but scaled version of the same solution pattern. The whole angular range can hence be described by repeated, scaled versions of a single-interval pattern. The other way round, the features identified for the pattern in one interval are valid for all other intervals as well. In the following we will employ the pattern between 13.2° and 21.8° for further analysis.

A distinct feature of the solution pattern is that it displays several features, which reflect a close relationship between mathematical number theory and the structural properties of Moiré crystals, which in turn may be critical for their physical properties. This is due to the fact that the discrete series of solutions along the continuous curves of eq.(19) (see Appendix Fig. A1) are interrupted at the limits of the intervals, i.e. at those angles, at which a lower-order solution exists. We can thus immediately understand that those red lines, the order of which is a number with many factors, contain a lower density of solutions. On the other hand, along those lines, the order of which is a prime number, the density of solutions is highest. This is a direct consequence of the fact that for orders corresponding to prime numbers no lower-order solutions exist, except $p$ = 1. Since in each interval defined by the presence of first- order solutions, $p$ solutions of $p^{th}$ order exist (see Appendix A1), we find sequences of $n_p - 1$ solutions along these lines, where $n_p$ is the respective prime number. The lines corresponding to an order, which is a square of a prime, have a number of gaps corresponding to that prime, and generally those corresponding to powers $r$ of a prime number $n_p$ have $n_p^{r-1}$ gaps.

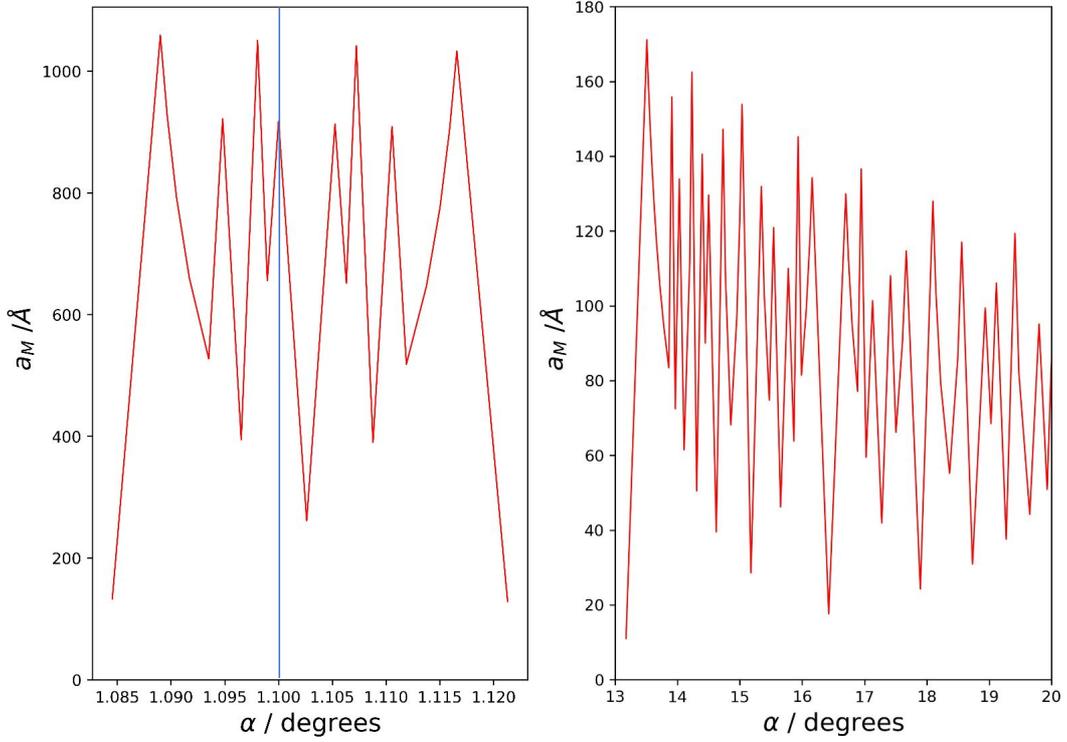

Fig. 7: Evolution of graphene Moiré crystal lattice parameters as a function of rotation angle, (a) around the magic angle 1.1° (blue line) taking into account orders up to 8, and (b) in the range from 13 to 20 degrees taking into account orders of up to 16.

For lines, the order of which does not correspond to a prime number, there are gaps in the sequence of solutions which correspond to solutions of their factors and solutions of the factors of their factors. For example, each interval for the line corresponding to the 6$^{th}$ order contains 2 solutions, since the two solutions being multiples of the two 3$^{rd}$ order solutions and the multiple of the 2$^{nd}$ order solutions are omitted. Lines corresponding to an order which has many factors, e.g. 10, 12, 24 etc., hence have particularly low densities of solutions. For a given order $p$, for all solutions which have a common divisor with $p$, a solution of lower order exists. Therefore, the number of solutions in each interval corresponds to the number of integers for which the greatest common divisor with $p$ is equal to 1. Or – in other words – the number of solutions in each interval is equal to the number of integers that are co-prime with $p$. The number of solutions in each interval as a function of the order $p$ is thus given by Euler's totient function $\varphi(p)$ (see Appendix Fig. A3). Indeed, if one counts the solutions on for each order, the characteristic sequence 1,1,2,2,4,2,6,4,6,4,10,… is found. The properties of the solution pattern are discussed in more detail in Appendix A2.2. Notably, in the solution pattern of the Moiré lattice parameters, which itself follows Euler's totient function, each solution defines a subset of solutions, which again follows Euler's totient function and thus establish a salient type of self-similarity. This is explained in more detail in Appendix A2.3.

**6. 30° case**

For the special case $\alpha = 30°$ the resulting pattern is not a Moiré lattice but a dodecagonal quasicrystal. The quasicrystal is nonperiodic and has thus no finite lattice parameter. This corresponds to the fact that no exact solution for a Moiré crystal lattice parameter exists for the 30° angle. If solutions for orders of up to 100 are calculated, the closest one deviates about $3 \cdot 10^{-3}$ % from 30° and has the order 97, and for the case of graphene would correspond to a lattice parameter of 8.7 μm. The lowest-order solution that falls well below a 1 % margin off 30° is of $26^{th}$ order and has a lattice parameter of 125 Å in graphene.

While the quasicrystal has no translation symmetry, it does possess a scaling symmetry, referred to as inflation symmetry [11]. This means, that the structure can be reproduced by scaling with a certain factor, i.e. it possesses a certain type of self-similarity. For the decagonal case the scaling factor is $\lambda = 2 + \sqrt{3}$. A second smaller scaling factor for this lattice exists, which leads to a reproduction of the structure when an additional rotation of 15° is taken into account. This second scaling factor is given by $\sqrt{\lambda}$.

The scaling factor $\lambda$ alternatively can be expressed as $1/\tan(30°/2)$. Therefore the index $m$ for the first-order Moiré point in the 30° case (12) equals $m_{30} = \lambda/\sqrt{3}$. With this value, we can calculate a hypothetical Moiré lattice parameter for the 30° case using (13) and obtain $a_{30} = \frac{1}{2}\sqrt{\lambda^2+1} = \sqrt{2+\sqrt{3}} = \sqrt{\lambda}$. The hypothetical lattice parameter of the Moiré lattice at 30° thus corresponds to the smallest scaling factor of the dodecagonal quasicrystal lattice.

**7. Discussion and conclusions**

In this paper a real-space approach for the calculation of lattice parameters of Moiré crystals forming by the relative rotation of two constituting hexagonal crystal layers is worked out, which can be applied to the case of graphene or metal-dichalcogenide crystals. A closed and consistent framework for the description of Moiré crystals and their structural parameters is provided, with solutions that are straightforward to implement. In the literature, predecessing papers including aspects of the present work are available. These, however, mainly focus on band-structure calculation and the interlayer electronic coupling [12,13] and restrict themselves to first-order Moiré crystals. The focus of the present paper, on the other hand, is rather on the discussion of the higher-order Moiré crystals and the structure of their solution pattern. One early paper, focusing on geometric aspects of Moiré crystals was published in 1993: Rong and Kuiper [14] carried out scanning tunneling microscopy (STM) images of [0 0 0 1] graphite surfaces. They observed a region with a top layer rotated by 2.1 degrees with respect to the bulk, and identified it as a Moiré superlattice with a period of 66 Å. Indeed this result corresponds to the clean Moiré point at 2.13° with a lattice parameter of 65.78 Å, listed in the fifth row of table 1 (scaled using a lattice parameter of 2.45 Å for graphite as in the reference). This pioneering paper also includes an expression for the continuous angle dependence of the first order Moiré points, which corresponds to eq.(14) in the present paper. In more recent STM work on twisted $WS_2$ bilayers [8] a relation between twist angle and Moiré period is established, which compares well to our results for first-order Moiré crystals. Higher-order Moiré lattices were considered by Lopes dos Santos, Peres, and Castro Neto [15], who calculate the Fourier components of the hopping amplitudes and show that in the low-angle limit only first-order solutions are relevant for the corresponding physical properties.

The results of the present paper imply that the angle dependence of the Moiré lattice parameter is critical. This directly relates to the required fabrication precision for the rotation angle of twin-layer devices. The current state-of-the-art for graphene bilayer systems is about 0.02° [16], and within this margin the resulting device may have considerably varying lattice parameters. If the exact value of the lattice parameter is critical for the functionality of the device, this may explain the low reproducibility of working devices, which is quoted as "…in three months of trying, just 2 of the 30 devices worked" [16]. On the other hand, the critical angle dependence allows for a highly accurate determination of the rotation angle in existing bilayer systems by means of transmission electron microscopy, via measurement of the lattice parameter, which can thus be used for quality control and reproducibility checks.

In summary, the present work allows for the following conclusions:

- The discussion of superstructures formed by rotated hexagonal structures requires discrimination between Moiré lattices, which apparently possess a continuous angle dependence, and Moiré crystals. For the latter, a basis is taken into account in addition to the constituting hexagonal lattice, and require a clear identification of "clean" Moiré points.
- For Moiré crystals, the local atomic order in a non-clean Moiré point can be significantly different from the reference point, i.e. the rotation center in the origin.
- The angle dependence of the Moiré-crystal lattice parameters has discrete solutions and higher-order Moiré points have to be taken into account. Solutions exist for all angles, but small changes of the rotation angle can lead to drastic changes of the resulting lattice parameter.
- For the upcoming field of twistronics, i.e. the deliberate fabrication of twisted bilayer devices with tailored physical properties, it is important to know the exact relation between the rotation angle and the structural parameters, in particular the lattice parameter, of the resulting devices.
- The angle dependence of the Moiré crystal lattice parameter has to be considered critical, which has consequences for the required fabrication precision for bilayer devices.
- The low reproducibility in the production of magic-angle devices may be a direct consequence of this critical angle dependence. It is highly recommendable to investigate functioning magic-angle devices by transmission electron microscopy to determine the actual angle.
- The solution pattern of Moiré crystal lattice parameters is highly complex and reflects a close relation between mathematical number theory and the formation parameters of the Moiré crystal, and thus potentially also the physical properties of twisted-bilayer devices.
- The formation of a twelvefold quasicrystalline structure at 30° corresponds to the case of a hypothetical first-order Moiré lattice parameter taking the value of the smallest inflation factor of the dodecagonal lattice.
- In the intervals defined by the positions of the first-order solutions the number of solutions as a function of the order $p$ follows Euler's totient function $\varphi(p)$.
- A novel type of self-similarity is found: in the solution pattern of the Moiré lattice parameters, which itself follows Euler's totient function, each solution defines a subset of solutions, which again follows Euler's totient function.

Finally it should be noted that the present treatment only considers the purely geometric case of rigid and non-interacting constituting lattices, in which any rotation of the lattice translates to exactly defined atom positions uniquely determined by the angle. In a real bilayer system relaxations may take place and locally lead to more favorable relative atom positions and thus an overall lower energy of the system. Recent scanning transmission electron microscopy investigations on homo- and heterogeneous bilayers of MoS2 and WS2 [17] and on MoSe2 and WSe2 heterostructures [18] indicate that atomic reconstruction indeed takes place. Local rearrangements may also lead to locking of favorable states under rotation and thus to a less continuous angle dependence than in the purely geometric case. In order to investigate such effects, furthergoing analysis e.g. by means of density-functional theory is required.

**Acknowledgments**

I am grateful to Dr. J.C. Fu for bringing me into contact with the topic of 2D superstructures, to Prof. R. Dunin-Borkowski for his encouragement and many inspiring discussions, and to Dr. J. Caron for careful reading of the manuscript.

**References**

[1] Y. Cao, V. Fatemi, S. Fang, K. Watanabe, T. Taniguchi, E. Kaxiras, P. Jarillo-Herrero (2018). Nature 556, 43 – 50.
[2] R. Bistritzer, A.H. MacDonald (2011). Proc. Natl Acad. Sci. USA 108, 12233 – 12237.
[3] E. Suárez Morell, J.D. Correa, P. Vargas, M. Pacheco, Z. Barticevic (2010) Phys. Rev. B 82, 121407.
[4] P. Moon, M. Koshino (2012). Phys. Rev. B 85, 195458.
[5] S. Fang, E. Kaxiras (2016). Phys. Rev. B 93, 235153.
[6] G. Trambly de Laissardiére, D. Mayou, L. Magaud (2012). Phys. Rev. B 86, 125413.
[7] A. Kerelsky, L.J. McGilly, D.M. Kennes, L. Xian, M. Yankowitz, S. Chen, K. Watanabe, T. Taniguchi, J. Hone, C. Dean, A. Rubio, A.N. Pasupathy (2019). Nature 572, 95 – 100.
[8] P.Y. Chen, X.Q. Zhang, Y.Y. Lai, E.C. Lin, C.A. Chen, S.Y. Guan, J.J. Chen, Z.H. Yang, Y.W. Tseng, S. Gwo, C.S. Chang, L.J. Chen, Y.H. Lee, (2019) Adv. Mater. 31, 1901077.
[9] X. Lu, X. Li, L. Yang (2019). Phys. Rev. B 100, 155416.
[10] N.R. Finney, M. Yankowitz, L. Muraleetharan, K. Watanabe, T. Taniguchi, C.R. Dean, J. Hone (2019). Nature Nanotechnol. 14, 1029 – 1034.
[11] J.E.S. Socolar (1989). Phys. Rev. B 39, 10519 – 10551.
[12] S. Shallcross, S. Sharma, O. A. Pankratov (2009). Phys. Rev. Lett. 101, 056803.
[13] S. Shallcross, S. Sharma, E. Kandelaki, O. A. Pankratov (2010). Phys. Rev. B 81, 165105.
[14] Z. Y. Rong and P. Kuiper (1993). Phys. Rev. B 50, 17427 – 17431.
[15] J.M.B. Lopes dos Santos, N.M.R. Peres, A.H. Castro Neto (2007). Phys. Rev. Lett. 99, 256802.
[16] https://physicstoday.scitation.org/do/10.1063/PT.6.1.20191119a/full/
[17] A. Weston, Y. Zou, V. Enaldiev, A. Summerfield, N. Clark, V. Zólyomi, A. Graham, C. Yelgel, S. Magorrian, M. Zhou, J. Zultak, D. Hopkinson, A. Barinov, T.H. Bointon, A. Kretinin, N.R. Wilson, P.H. Beton, V.I. Fal'ko, S.J. Haigh, R. Gorbachev (2020). Nature Nanotechnol. https://doi.org/10.1038/s41565-020-0682-9
[18] M.R. Rosenberger, H.J. Chuang, M. Phillips, V.P. Oleshko, K.M. McCreary, S.V. Sivaram, C. S. Hellberg, B.T. Jonker (2020). ACS Nano 14, 4550 – 4558

# Appendix

## A1: Structure of the diagram of higher-order Moiré points

The angle dependence of the higher-order Moiré lattice vectors calculated according to (19) and the length of the vectors pointing at higher-order Moiré points are shown in Fig. A1. This figure corresponds to the (differently scaled) figure 3a in [15], where the first 6 orders are displayed.

Figure A1 is the basis for the determination of the lattice parameters of Moiré crystals, for which at each angle only the solution of lowest order is regarded.

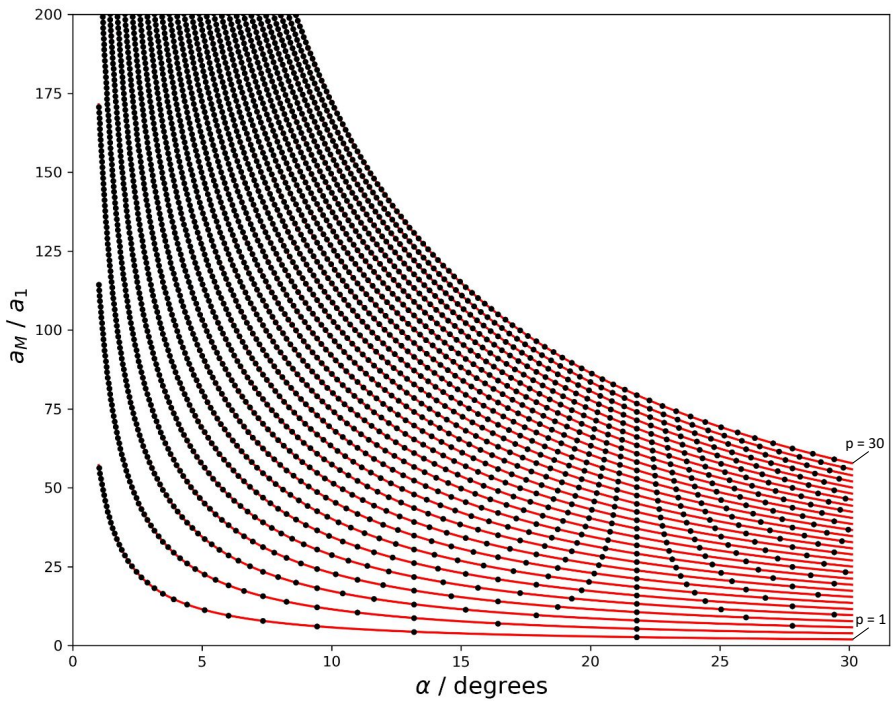

Fig. A1: Length of higher-order Moiré lattice vectors (in units of the reference lattice parameter) as a function of the rotation angle $\alpha$ for $p$ = 1 to 30; black: length of actual higher-order Moiré-lattice vectors for which $m$ and $n$ are integer.

The general structure of the solution pattern in Fig. A1 can be understood as follows: the first-order solutions are solutions for all higher orders as well, because the lattice parameters of the higher-order patterns are multiples of those of the first order. These sets of solutions make up the vertical sets of points, dividing the $\alpha$-axis into intervals. The interval from 13 to 20 degrees is shown in Fig. A2, and similar intervals, becoming narrower at lower and wider at higher angles, exist between all first-order solutions. Within each interval, there exists one solution for the first order, two solutions for the second order, three for the third, and so forth. Since all solutions correspond to periodicities of Moiré crystals, all second-order solutions are solutions for the 4$^{th}$ order, 6$^{th}$ order, etc. Generally all $p^{th}$ order solutions are solutions of the $n \cdot p^{th}$ order as well (blue arrows in Fig. A2).

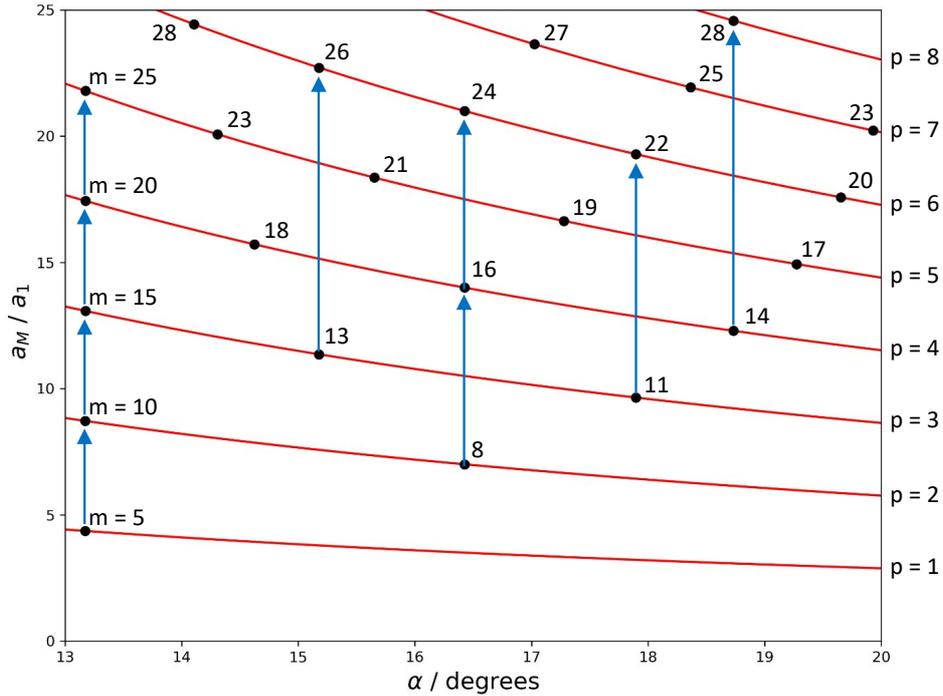

Fig. A2: Section of the pattern in Fig A1, displaying the index *m* for each point and arrows indicating the relation between lower-order points and their higher-order multiples.

These rules define the general structure of the pattern. The positions of the points on the $\alpha$-axis are defined by their index *m* according to (17). As shown above, the angle dependence can be approximated by the inverse of the angle, so relative positions $\alpha_i$ and $\alpha_j$ are determined by the inverse ratio of their respective index *m*, i.e. $\alpha_i / \alpha_j = m_j / m_i$.

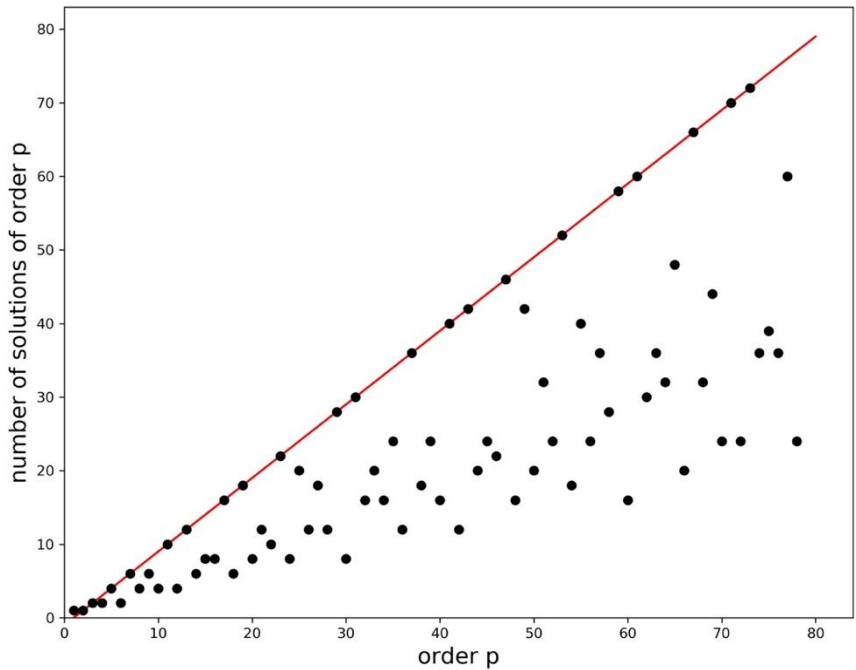

Fig. A3: The number of solutions of a given order as a function of the order *p*, within an interval between two first-order solutions. The line represents the prime numbers. The number of solutions is described by Euler's totient function $\varphi(p)$.

## A2: Structure of the solution pattern for Moiré crystals

*A2.1: Euler's totient function*

The solution pattern for Moiré crystals is subdivided into intervals defined by the angles at which first-order solutions exist. It is argued in the text that within each of these intervals the number of solutions is given by Euler's totient function $\varphi(p)$. This is illustrated in Fig. A3, displaying the number of solutions in the interval from 13.2 to 21.8 degrees as a function of the order $p$ for the first 80 orders in the typical representation of $\varphi(p)$. The red line represents the prime numbers $n_p$ for which $\varphi(n_p) = n_p - 1$.

*A2.2: Pole functions*

In Fig. 6 it can be seen that at the angle of each first-order solution, an apparent pole exists, at which the angle dependence of the higher-order solutions diverges. This is very obvious at e.g. 9.4°, 13.2° and 21.8°, but at lower angles narrower poles also exist for each first-order solution. We will refer to the angle dependence of the sequences of solutions forming these apparent poles as pole functions. They can be analytically described in a very similar way as the continuous description of the Moiré lattice vectors (19) and its approximation by a hyperbolic function. In particular, the pole functions can be described as

$$f_{n,p}(\alpha) = \left| n \cdot \frac{\sqrt{3}}{p} \cdot \frac{\alpha_p}{\alpha_p - \alpha} \right|, \quad (20)$$

where $\alpha_p$ is the angle at which a solution of order $p$ exists and $n$ is an integer. The equation thus does not only describe the pole functions corresponding to first-order solutions, but the more general family of curves corresponding to any solution.

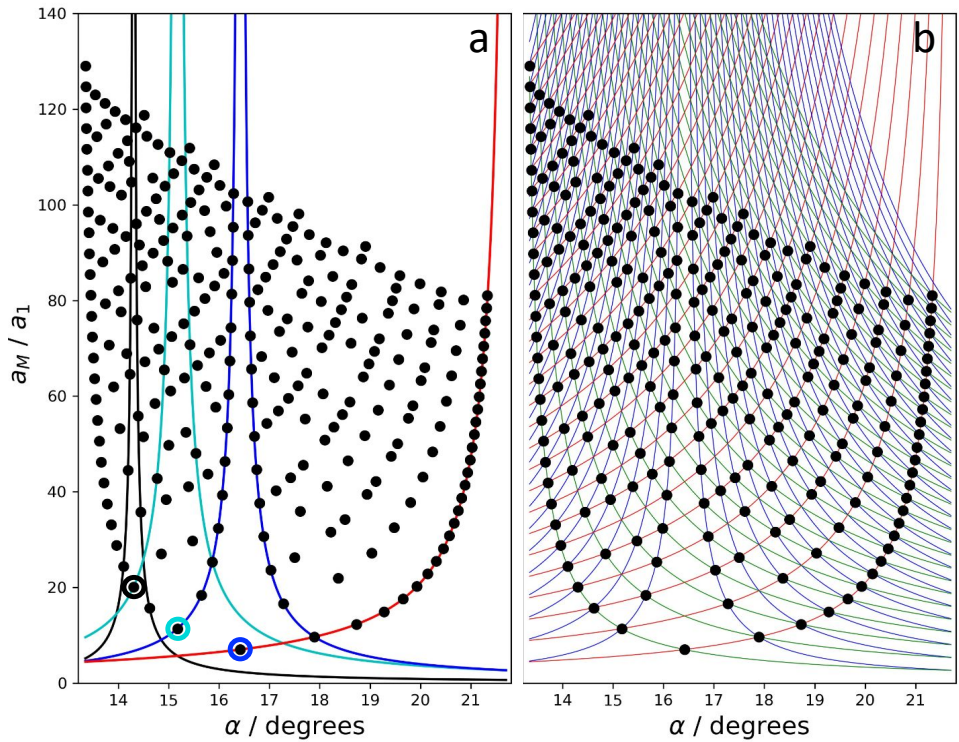

Fig. A4: Pole functions for the Moiré crystal solution pattern in the interval from 13.2 to 21.8 degrees. (a) Examples red: *p* = 1, *n* = 1; *blue: p* = 2, *n* = 1; cyan: *p* = 3, *n* = 2; black: *p* = 5, *n* = 1. (b) Family of pole functions for *p* = 1 (red and green), *p* = 2 (blue) and *n* = 1 to 30.

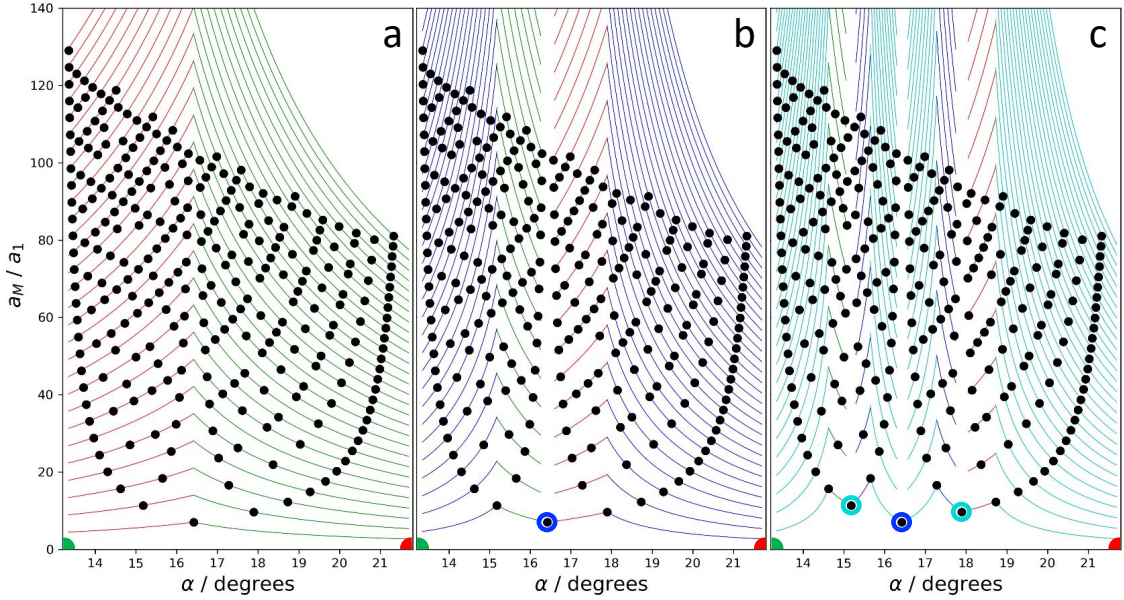

Fig. A5: Self-similarity and Euler's totient function (see text).

The red line in Fig. A4a represents the pole function calculated according to (20) formed at the first-order solution at $\alpha_p$ = 21.8° and for the case *n* = 1. The blue and black lines represent the pole function for the second-order solution at 16.4° and the fifth-order solution at 14.3°, respectively, both for n = 1. The cyan lines represent the pole function for n = 2 of the third-order solution at 15.2°. The solutions to which the pole functions correspond are marked with circles of the same color.
In Fig. A4b the families of pole functions for *n* = 1 to 30 corresponding to the first-order solutions at 13.2° (green) and 21.8° (red), and the second-order solution at 16.4° (blue), are shown. It is obvious that the solutions are all located on crossing points of the curves, but not all crossing points entail a solution. This would be the case, if not solutions for Moiré crystals (c.f. Fig. 6), but solutions for Moiré lattice vectors (c.f. Fig. A1) were considered.

*A2.3: Self-similarity*
For each family of pole functions the sequence of solutions on a given pole curve reflects the divisibility of the index *n* of the family member. Again, Euler's totient function is hidden in this graph, which can be seen in Fig. A5a: if you count the number of solutions in the subdivision $\propto\ \leq\ 16.4°$ for each *n* on the pole functions corresponding to the first-order solution at 13.2° (red lines and point), you obtain the sequence 1,1,2,2,4,2,6,4,6,4,10,…, i.e. the number of solutions follows Euler's totient function $\varphi(n)$. In the same graph, the same is found for $\propto\ \geq\ 16.4°$ for the pole functions corresponding to the first-order solution at 21.8° (green lines and point). This recurrence of Euler's totient function is also found for higher-order solutions: Fig A5b shows the family of pole functions corresponding to the second-order solution at 16.43° (blue lines and circle), where the number of solutions in the subdivisions $\propto\ \leq\ 15.2°$ and $\propto\ \geq\ 17.9°$ is given by $\varphi(n)$. The remaining solutions in the subdivisions $15.2°\ \leq\ \propto\ <\ 16.4°$ and $16.4°\ <\ \propto\ \leq\ 17.9°$ fall on the family of pole functions corresponding to the first-order solutions (red and green), and for both individually the number of solutions is given by $\varphi(n)$. Fig. A5c shows the same for the third-order solutions at 15.2° and 17.9° (cyan lines and circles). In the four corresponding subdivisions, the solutions individually follow $\varphi(n)$, and the remaining subdivisions are covered by lower-order solutions (here second- and first-order), each individually following $\varphi(n)$. This can be continued *ad-infinitum* for each solution in the pattern – each solution defines a family of pole functions and a set of angular subdivisions, in each of which the solutions follow $\varphi(n)$, and the rest of the pattern is subdivided into ranges covered by solutions of the same and lower orders, for which the same holds.

We thus find a novel type of self-similarity in the solution pattern of the Moiré lattice parameters: in each interval defined by the poles at the first-order solutions the number of solutions for each order *p* is given by $\varphi(p)$, and each of those solutions subdivides the angular interval into further sub-intervals, in which the number of solutions for a given *n* of the corresponding pole-function family again is given by an Euler totient function $\varphi(n)$.